\documentclass[aps,prb,groupedaddress,showpacs,nofootinbib]{revtex4}
\usepackage{amsmath,amscd,amssymb,epsfig}
\usepackage{amsmath}
\usepackage{graphicx}
\begin{document}
\bibliographystyle{apsrev}
\title[Subdiffusive dynamics in washboard potentials]
{Subdiffusive dynamics in washboard potentials:
two different approaches \\and different universality classes \label{xxx}}

\author{Igor Goychuk}
\author{Peter H\"anggi}
%\footnote{Author footnote.}

%\index[aindx]{Author, F.} % or \aindx{Author, F.}
%\index[aindx]{Author, S.} % or \aindx{Author, S.}

\affiliation{Institute of Physics, University of Augsburg, \\
Universit\"atsstr. 1, D-86135 Augsburg, Germany, \\
goychuk@physik.uni-augsburg.de %\footnote{Affiliation footnote.}
}

\begin{abstract}
We consider and compare two different approaches to the fractional subdiffusion
and transport in washboard potentials. One is based on the concept of random 
fractal time and is associated with the fractional Fokker-Planck equation. 
Another approach is based
on the fractional generalized Langevin dynamics and is associated with anti-persistent 
fractional Brownian motion and its generalizations.
Profound differences between these two different approaches sharing 
the common adjective ``fractional'' are explained
in spite of some similarities they share in the absence of a nonlinear force.
In particular, we show that the asymptotic dynamics in tilted washboard potentials
obey two different universality classes independently of the form of potential.

\end{abstract}

\maketitle

\section{Introduction}\label{sec1.1}

\begin{figure}
\vspace{1cm}
\includegraphics [width=0.5\linewidth]{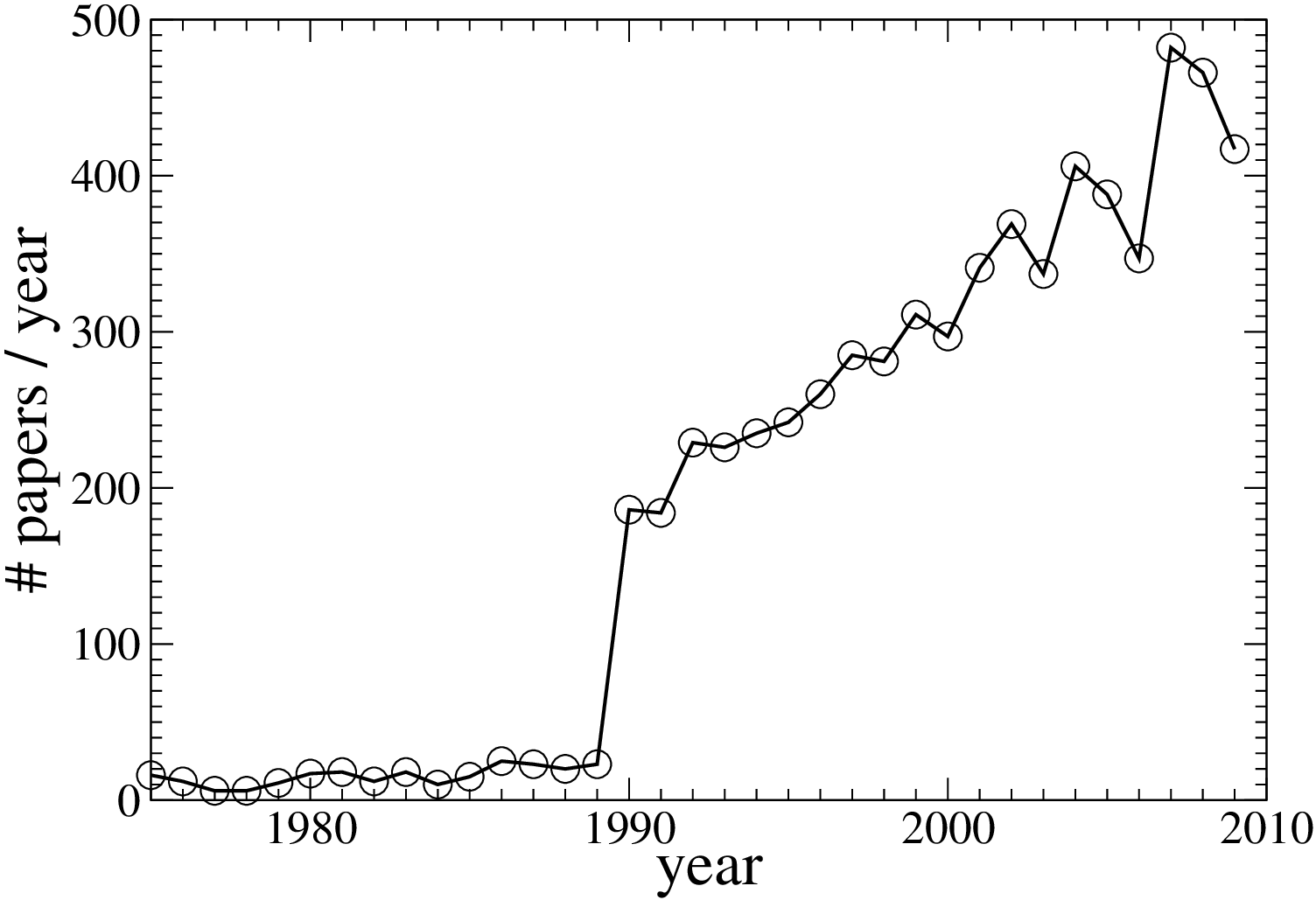}
\caption{Number of papers on the subject ``anomalous diffusion'' published
per year according to the ISI Web of Science, Thomson-Reuters. 
Notice the ``phase transition''-like rise during 1990.}
\label{Fig1}
\end{figure}

Anomalous diffusion becomes an increasingly popular subject with the number of papers
published per year growing fast over last twenty years after a distinct
rise occurred in 1990 
(cf. Fig. \ref{Fig1}). Since then, 
it spreads  from such typical physical applications as
charge transport in disordered solids and hot plasmas to biophysical
applications and even quantitative finance 
\cite{Shlesinger1,Scher,Bouchaud,Hughes,Shlesinger,
Metzler,West,West2,Zaslavsky,Saxton,Seisenberg,Tolic,GH04,Golding,Wilhelm,
Amblard,Szymanski,Kou,Mizuno,Reuveni}. 

There are several sufficiently generic 
physical mechanisms and accompanying theoretical approaches to describe the complexity
of anomalous transport processes. One approach is intrinsically based on the physical
picture of a stochastic time clock 
\cite{Shlesinger1,Scher,Hughes,Shlesinger,Metzler,Sokolov1}. It models random 
sojourns of a travelling
particle in trapping domains of a disordered solid, e.g. due to energy disorder.
After a random time spent in some spatially located trapping domain the particle jumps 
to a neighboring domain, or maybe farther, and such a jumping process
continues in time. The jump directions and their
lengths are not correlated from jump to jump and the next clock period
is not correlated with those passed (semi-Markovian assumption 
\footnote{This does not exclude infinite range memory
leading to a weak ergodicity breaking \cite{Barkai}.}). Such a random
clock is completely characterized by the probability density of clock periods 
$\psi(\tau)$.
In a given time interval $t$ there will
be a random number of jumps $n$, or 
stochastic clock 
periods completed. However, if the mean clock 
period $\langle \tau\rangle$ exists,
the probability distribution $p(n,t)$ of ``ticking'' $n$ times within the 
observer time window $t$  becomes for large $n$ a very sharp function \cite{Hughes} around 
$n^*(t)=t/\langle \tau\rangle$, as characterized by the 
relative dispersion, $\langle \delta n^2(t) \rangle^{1/2}/
\langle n(t) \rangle$ (see Appendix A). 
In the continuum medium 
approximation the trapping domains shrink to points. 
Then, $\langle \tau \rangle$ can be made arbitrarily small and $n$ becomes 
quasi-continuous variable for a finite $t$.
Correspondingly, the probability density of intrinsic time, 
$\tau(t)=n(t)\tau_{\rm sc}$,
where $\tau_{\rm sc}$ is a time-scaling parameter, an intrinsic clock 
time unit equal to the
duration of time period for the regular clock, 
assumes a delta-function,
$p(\tau)=\delta(\tau-t)$, and the stochastic clock is not different
from the regular one.

The situation changes dramatically if the mean of
sojourns does not exist, or better to say, it exceeds largely a 
typical time required to
diffuse across  the physical medium of a finite size. 
Then, $p(n,t)$ is not a sharp
function around the mean number $\langle n(t)\rangle$.
This is the case, for example, when $\psi(\tau)$ possesses
a long tail, $\psi(\tau)\propto 
(\tau/\tau_{\rm sc})^{-1-\alpha}$ for $\tau/\tau_{\rm sc}
\to \infty$ with $0<\alpha<1$.
This implies a divergent mean $\langle \tau\rangle \to\infty$, and
diverging higher moments as well.
Nevertheless,  $\langle n(t)\rangle \propto (t/\tau_{\rm sc})^{\alpha}$ 
exists for any finite $t$ and it scales 
sublinearly with the physical time $t$ (see in Appendix A). In terms of 
the one-sided Levy distribution density ${\cal L}_{\alpha}(z)$, 
$p(n,t)=(t/\tau_{\rm sc}){\cal L}_{\alpha}(n^{-1/\alpha}
t/\tau_{\rm sc})/(\alpha n^{1/\alpha+1})$. 

Consider now an ensemble of particles.
Until time $t$, each particle 
has accomplished an {\it individual} number of intrinsic time periods corresponding
to the intrinsic time $\tau(t)$ which becomes a random
variable broadly distributed: all the particles have their own history, 
maintaining individuality 
and avoiding the fate of self-averaging even in the strict limit $t\to\infty$. 
Only an additional ensemble averaging
smears out this principal randomness\cite{Sokolov1,He,Lubelski}. 
The {\it unbiased} diffusion becomes anomalously slow and nonergodic with the spatial 
variance of a {\it cloud} of particles 
growing sublinearly, 
$\langle \delta x^2(t)\rangle \propto \langle n(t)\rangle \propto 
t^{\alpha}$. 
Such a nonergodic approach
to subdiffusion seems appropriate for disordered solids, e.g. thin amorphous 
films \cite{Scher,Hughes}, with a more recent example 
provided by ${\rm Ti O_2}$ nanocrystalline electrodes in 
the Gr\"atzel's
photovoltaic cell elements \cite{Nelson}. For charged carriers within 
such media one can create
a potential energy profile $U(x)$ by applying an electrical field of spatially 
distributed fixed charges and a static external electrical field. Then in the continuum
approximation subdiffusion can be described by the fractional Fokker-Planck
equation (FFPE) \cite{Metzler,Metzler99,Barkai01}
\begin{eqnarray}\label{FFPE1}
 \frac{\partial}{\partial t} P(x,t) = \sideset{_0}{_t}{\mathop{\hat
D}^{1-\alpha}} \left [- \frac{\partial}{\partial x} 
\frac{f(x)}{\eta _\alpha} + \kappa _\alpha \frac{\partial ^2}{\partial x^2}
\right ] P(x, t) \,, 
\end{eqnarray}
where $f(x)=-d U(x)/dx$ is the force,  
$\eta _\alpha$ is the fractional friction coefficient related to the
fractional subdiffusion coefficient $\kappa_{\alpha}$ by the generalized Einstein
relation, $\eta_{\alpha}=k_BT/\kappa_{\alpha}$, at temperature $T$ and 
\begin{equation} \label{RL}
\sideset{_{t_0}}{_t}{\mathop{\hat D}^{\gamma}} P (x, t) =\frac{1}{
\Gamma(1-\gamma)} \frac{\partial}{\partial t} \int_{t_0}^{t}
\mathrm{d} t' \, \frac{P(x, t')}{(t-t')^{\gamma}} \, ,
\end{equation}
is the Riemann-Liouville operator of the fractional derivative 
\cite{Mainardi}, where $0<\gamma<1$ and $\Gamma(x)$ is the gamma-function.
The FFPE (\ref{FFPE1}) can be derived within the above continuous time random walk 
(CTRW) framework. It can
also be written in the form using the 
Caputo fractional derivative
\cite{Mainardi}
\begin{eqnarray}
\sideset{_{t_0}}{_{*}}{\mathop{D}^{\gamma}}P(x,t):=\frac{1}{\Gamma(1-\gamma)}
\int_{t_0}^t dt' \frac{\partial P(x,t')/\partial t'}{(t-t')^\gamma}
\end{eqnarray}
acting on the left hand side, yielding~\cite{GH06}
\begin{eqnarray}\label{FFPE2}
\sideset{_{0}}{_{*}}{\mathop{D}^{\alpha}}P(x,t)= 
\kappa_{\alpha} \frac{\partial }{\partial
x} \left ( e^{-\beta U(x)} \frac{\partial}{\partial x} \, e^{\beta
U(x)} P(x,t)  \right )=-\frac{\partial J(x,t)}{\partial x} \,,
\end{eqnarray}
in the transport form.
Here, $\beta=1/(k_BT)$ is inverse temperature and 
\begin{eqnarray}
J(x,t)=-\kappa_{\alpha}
e^{-\beta U(x)} \frac{\partial}{\partial x} \, e^{\beta
U(x)} P(x,t)
\end{eqnarray}
is the subdiffusive flux.
 It should be emphasized that a non-Markovian Fokker-Planck equation  never defines
the corresponding non-Markovian process completely \cite{HT77,Grabert}. 
It allows to find merely the 
single-time, conditional, and double-time probability densities, but never the 
multi-time probability densities.
However, the FFPE dynamics can be nicely 
simulated from the underlying CTRW with the nearest neighbors jumps only 
\cite{GH06,H06}.   
 
A quite different approach to subdiffusion is associated with 
the fractional Brownian motion \cite{Mandelbrot}.
Here, the principal issue is the long-range anticorrelations in the particle 
displacements, positive increments follow with a greater probability by negative 
increments and {\it vice versa}\cite{Goychuk09,Jeon}, which can reflect e.g. the phenomenon of viscoelasticity in complex 
glass-forming liquids above, but close to the glass transition. The Brownian 
particle
is temporally trapped in a trap (cage effect) by an elastic force with spring constant
$G(t)$ which decays to zero
in time releasing the particle. Let's assume that the motion starts
at $t_0$, $v(t)=\dot x(t)=0$ for $t<t_0$. 
In the linear approximation, on the particle  acts a 
viscoelastic force $F_{\rm v-el}(t)=-\int_{t_0}^t G(t-t')
\dot x(t')dt'$, where $\dot x(t)$ is the particle's instant velocity. 
The first theory of 
viscoelasticity has been proposed by J. Clerk Maxwell \cite{Maxwell} in 1867. 
It corresponds to $G(t)$ exponentially decaying
in time, $G(t)=G_0\exp(-t/\tau_0)$, with a relaxation time constant $\tau_0$. 
Departing from the phenomenon of elasticity in solids 
Maxwell derived the phenomenon of viscosity in liquids
in the limit where the decay of elastic modulus is very fast on the time scale of
$v(t)$ change, which corresponds to 
$G(t)=2\eta_0 \delta(t)$, with $\eta_0=G_0\tau_0$ 
being the viscous friction coefficient. 
In the theory of generalized Brownian motion, $G(t)$
is interpreted as the frictional memory kernel $\eta(t)$, rather than
a decaying elastic force constant. Here, one departs from
the phenomenon of viscosity and viscous Stokes memoryless friction and considers 
the emerging 
elasticity in complex fluids or viscoelastic bodies. Both view points are
essentially equivalent for a positive $\eta(t)>0$ departing just from different 
standing points \footnote{$\eta(t)$ can also be negative, e.g. accounting for
the hydrodynamic memory or in the case
of superdiffusion. Therefore, the memory-friction interpretation is, 
in fact, more general.}. 
The anticorrelations in the particle's displacements are due to the
elastic restoring force component. 

In complex media, the memory function $G(t)$ is better described
by a sum of exponentials reflecting a viscoelastic response with multiple time scales. 
Moreover, 
in 1936 A. Gemant \cite{Gemant} found that some viscoelastic
bodies are better described by a $G(t)$ relaxing in accordance with a
power law, $G(t)\propto t^{-\alpha}$,  rather than a single-exponential 
and introduced
a fractional integro-differential in the viscoelasticity theory. Using the notion of
fractional Caputo derivative 
such a visco-elastic force can be short-handed, written as 
\begin{equation}\label{model2}
 F_{\rm v-el}(t)=-\eta_{\alpha}
 \sideset{_{t_0}}{_{*}}{\mathop{D}^{\alpha}} x(t)\;.
\end{equation}
Indeed, such and similar viscoelastic responses are measured \cite{Wilhelm,Mizuno,Mason} 
using the microrheology
methods  \cite{rheo}. The Brownian motion never stops and the frictional 
loss of energy must be compensated on average by the energy gain
provided by a zero-mean random force of environment so that at the thermal equilibrium 
the equipartition theorem holds, in accordance with
the classical fluctuation-dissipation theorem. 
Within the considered model of a linear 
memory-friction such a force must be Gaussian \cite{Reimann} 
(but not necessarily so beyond the linear friction model). 
As a result, the Brownian motion
of a particle of mass $m$ is described by the Fractional Langevin Equation (FLE)
\cite{Lutz,Coffey,Goychuk07a,Burov} 
\begin{equation}\label{fle}
m\ddot x+ \eta_{\alpha}
\sideset{_{0}}{_{*}}{\mathop{D}^{\alpha}} x(t)=f(x)+\xi(t) \;,
\end{equation}
(from now on we fix $t_0=0$) which is a particular case of the celebrated 
Generalized Langevin Equation (GLE) \cite{Kubo1,Kubo2,Zwanzig,HTB90,WeissBook}
\begin{equation}\label{gle}
m\ddot x+\int_{0}^{t}\eta(t-t')\dot x(t')dt'=
f(x)+\xi(t) \;,
\end{equation}
with the memory kernel $\eta(t)=\eta_{\alpha}t^{-\alpha}/\Gamma(1-\alpha)$ 
and the noise autocorrelation 
function obeying the fluctuation
dissipation relation 
\begin{equation}\label{fdt} \langle
\xi(t)\xi(t')\rangle=k_BT\eta(|t-t'|) \;.
\end{equation}
Such a GLE can be derived also from a Hamiltonian model for a 
particle bilinearly 
coupled with coupling constants $c_i$
to a thermal bath of harmonic oscillators with masses $m_{i}$ and frequencies
$\omega_i$, $H_{B,\rm int}(p_i,q_i,x)=(1/2)\sum_i \{p_i^2/m_i+m_i\omega_i^2
[q_i-c_i x/(m_i\omega_i^2)]^2\}$.
The total effect of the bath oscillators,
which are initially canonically distributed with $H_{B,\rm int}$ 
at temperature $T$ and fixed $x=x(0)$, 
is characterized
by the bath spectral density
\begin{equation}
J(\omega) = \frac{\pi}{2} \sum_i
\frac{c_i^2}{m_{i} \omega_{i}} \delta(\omega-\omega_i).
\label{J}
\end{equation}
The memory kernel is $\eta(t)=(2/\pi)\int_0^{\infty}J(\omega)\cos(\omega t)d\omega$ 
in terms of $J(\omega)$ and the subdiffusive FLE corresponds to a 
sub-Ohmic, or fracton thermal bath 
with $J(\omega)=\eta_{\alpha}\sin(\pi\alpha/2)
\omega^{\alpha}$ \cite{WeissBook}. 
Without frequency cutoffs such a model presents a 
clear idealization.
There always exists a highest frequency of the thermal bath and this leads to 
a small time  regularization of the memory kernel, i.e. a short-time cutoff. 
Physically, this takes into account the medium's granularity beyond the continuum
approximation. Moreover, in the case of a finite-size medium 
there always exists also a smallest frequency of the medium's oscillators corresponding
to the inverse size of the medium. These facts become especially clear if one evaluates
the spectral density of low-frequency ``fracton'' oscillators in proteins, 
see in Refs. \cite{Reuveni}.
This leads also to a cutoff at large times 
in the memory kernel and the dynamics can be
subdiffusive on the time scale smaller than the corresponding memory cutoff.
The latter one can be but prominently large which makes the considered 
idealization relevant. Important is also the result that an overdamped
FLE description  of subdiffusion can be derived 
from a broad class of phenomenological continuum elastic 
models \cite{Taloni}.

In the inertialess limit with $m\to 0$, 
one can conceive the idea that FFPE (\ref{FFPE1},
\ref{FFPE2})
is the fractional Fokker-Planck equation corresponding to  the FLE (\ref{fle}). 
This idea is but wrong \cite{Goychuk07a}. 
The non-Markovian Fokker-Planck equation (NMFPE) which corresponds
to the GLE (with arbitrary kernel) \cite{Adelman,Hanggi78,Hynes} and 
to the FLE, in particular\cite{Goychuk07a},  
is a different one. Presently, its explicit form
is known only for constant or
linear forces $f(x)$\cite{Adelman,Hanggi78,Hynes}. 
This resulting NMFPE 
has the form of 
Fokker-Planck equation with 
time-dependent kinetic coefficients. This time-dependence
is not universal and it heavily depends on the form of potential. In turn, 
the Langevin equation which corresponds to the above FFPE is known and it has 
the form of a Langevin equation which is local in the stochastic time $\tau(t)$
and describes thus a doubly stochastic process \cite{DoublyStoch}.
Here lies also the profound mathematical difference between these two approaches
to subdiffusion. The physical differences are also immense. In particular, the 
GLE and FLE
approaches are asymptotically mostly ergodic
as they are not based on the concept of 
fractal stochastic time
with divergent mean period and the mean residence time in a finite spatial
domain remains finite. 
Before we discuss the striking differences in more detail, 
let us start from some apparent, but misleading similarities.

\section{Free subdiffusion and constant bias}

Free subdiffusion, as well as diffusion biased by a constant force $F$ can readily be 
solved in both approaches using the method 
of Laplace-transform. First one finds the Laplace-transform of the mean 
ensemble-averaged displacement $\langle \delta x(t)\rangle $, and of the position 
variance
$\langle \delta x^2(t)\rangle= \langle x^2(t)\rangle-
\langle x(t)\rangle^2$, starting from a delta-peaked distribution at $x=0$ and 
$t_0=0$. Then, one transforms back to the time domain.  This gives\cite{Metzler} 
\begin{eqnarray} \label{mean}
\langle \delta x(t)\rangle=\mu_{\alpha} F t^{\alpha}/\Gamma(1+\alpha)
\end{eqnarray}
and %for $F=0$
\begin{eqnarray} \label{variance}
\langle \delta x^2(t)\rangle=2\kappa_{\alpha}t^{\alpha}/\Gamma(1+\alpha)
\end{eqnarray}
with the generalized mobility $\mu_{\alpha}=1/\eta_{\alpha}$
related to the subdiffusion coefficient at $F=0$ by the generalized Einstein 
relation $\mu_{\alpha}=\kappa_{\alpha}/(k_BT)$. 
Within the FLE approach these results are valid in the strict inertialess 
limit $m\to 0$. Furthermore, the Eq. (\ref{variance}) is still valid  then for 
{\it arbitrary} $F\neq 0$. However, within the FFPE approach the Eq. (\ref{variance})
is valid {\it only} for $F=0$, which is the first striking difference, see also below. 
Furthermore, both results are also valid
asymptotically, $t\to\infty$, within the FLE  for a finite $m\neq 0$.

 Generally, the GLE results
can be obtained for arbitrary memory kernel $\eta(t)$. Assuming the particles
being initially Maxwellian distributed, i.e. thermalized with
thermal r.m.s. velocities 
\begin{eqnarray}
v_T=\sqrt{k_BT/m}, 
\end{eqnarray}
one can
obtain for the Laplace-transformed stationary velocity (fluctuation) 
autocorrelation function (VACF)
$K_v(\tau)=\langle \delta v(t+\tau)\delta v(t)\rangle$, 
$\delta v(t)=v(t)-\langle v(t)\rangle$,
\begin{eqnarray}
\tilde K_v(s)=\frac{k_BT}{ms+\tilde \eta(s)},
\end{eqnarray}  
where $\tilde \eta(s)$ is the Laplace-transform of $\eta(t)$.
This is a well-known result which was obtained first by Kubo\cite{Kubo1,Kubo2} 
in the Fourier space.
For the FLE with $\tilde \eta(s)=\eta_{\alpha}s^{\alpha-1}$ it 
yields by the inversion to the time-domain \cite{Lutz}
\begin{eqnarray}\label{vfle}
K_v(\tau)=v_T^2 E_{2-\alpha}
[-(\tau/\tau_{v})^{2-\alpha}]
\end{eqnarray}
with $\tau_{v}=(m/\eta_{\alpha})^{1/(2-\alpha)}$ being the anomalous 
velocity relaxation time constant. In (\ref{vfle}), 
$E_{\gamma}(z)$ is the Mittag-Leffler
function, $E_{\gamma}(z)=\sum_{n=0}^{\infty}z^n/\Gamma(n\gamma+1)$ \cite{Metzler}.
For $0<\alpha<1$, $K_v(\tau)$ is initially positive reflecting 
ballistic persistence due to
inertial effects
and then becomes negative (anti-persistence due to decaying elastic cage force). 
In the limit $m\to 0$, the VACF undergoes a jump starting from 
$v_T^2$ at $\tau=0$ and
then becoming negative, 
$K_v(\tau)\propto -1/\tau^{2-\alpha}$ for $\tau>0$, corresponding
to purely anti-persistent motion. The position variance is given by the 
doubly-integrated VACF. Its Laplace-transform therefore reads,
\begin{eqnarray}\label{x2}
\langle \widetilde{\delta x^2(s)}\rangle =\frac{2k_BT}{s^2[ms+\tilde \eta(s)]}.
\end{eqnarray}
Moreover, 
\begin{eqnarray}
\langle \widetilde{\delta x(s)}\rangle =\frac{F}{s^2[ms+\tilde \eta(s)]},
\end{eqnarray}
for arbitrary kernel, which can also be easily shown from the GLE,
 and therefore \footnote{For nonequilibrium
initial preparations this result holds asymptotically in any asymptotic 
ergodic case, 
including the FLE dynamics. 
The relaxation to the asymptotic regime, or aging, can be but very
slow \cite{Pottier,Deng} which is a general feature of subdiffusive GLE dynamics 
also in periodic potentials.} 
\begin{eqnarray}\label{normal}
\frac{\langle \delta x(t) \rangle}{
\langle \delta x^2(t) \rangle}=\frac{F}{2k_BT} \;
\end{eqnarray}
for the thermally equilibrium initial preparation.
For the FLE with a finite $m$ the inversion of Eq. (\ref{x2}) to the time domain
yields\cite{Lutz}, 
\begin{eqnarray}
\langle \delta x^2(t) \rangle=2v_T^2t^2 E_{2-\alpha,3}
[-(t/\tau_{v})^{2-\alpha}], 
\end{eqnarray}
where $E_{\gamma,\beta}(z)=
\sum_{n=0}^{\infty}z^n/\Gamma(n\gamma+\beta)$ is the generalized Mittag-Leffler
function. One recovers Eq. (\ref{variance}) in the limit $m\to 0$.

However, for the subdiffusive CTRW
and FFPE dynamics the behavior of the ensemble-averaged variance is very different 
from Eq. (\ref{variance}) under a non-zero bias
$F\neq 0$. Then, the Eq. (\ref{variance}) is not valid anymore. 
This fact is ultimately related to the properties of the stochastic
clock. The point is that starting from a CTRW picture it is easy 
to show (see Appendix A) 
that the growing ensemble-averaged 
variance $\langle \delta x^2(t)\rangle$ depends in the asymmetric case
(the probabilities to jump left and right are different) not only on the
mean number $\langle n(t)\rangle$ of the stochastic clock periods passed, 
but also on their variance $\langle \delta n^2(t)\rangle$.  For $\alpha=1$
(regular clock), $\langle \delta n^2(t)\rangle$=0. However, for
$0<\alpha<1$, $\langle \delta n^2(t)\rangle \propto t^{2\alpha}$ and this
dramatically changes the character of anomalous CTRW and FFPE diffusion in the
presence of bias. It becomes asymptotically 
$\langle \delta x^2(t)\rangle\propto F^2 t^{2\alpha}$, while 
$\langle \delta x(t)\rangle\propto F t^{\alpha}$. Notice that for $1/2<\alpha<1$
the subdiffusion at $F=0$ transforms into superdiffusion for $F\neq 0$, i.e. a cloud
of particles spreads out anomalously fast relative to its center of mass. 
This yields a remarkable
scaling for the ensemble-averaged quantities
\begin{eqnarray} \label{anomal}
\lim_{t \to \infty} \frac{\langle \delta x^2(t) \rangle}{
\langle \delta x(t) \rangle^2} = 
\lim_{t \to \infty} \frac{\langle \delta n^2(t) \rangle}{
\langle n(t) \rangle^2}=\frac{2
\Gamma ^2(\alpha + 1)}{\Gamma(2 \alpha + 1)} - 1 \, .
\end{eqnarray}
This scaling, which was observed first in Refs. \cite{Shlesinger1,Scher} for a CTRW
subdiffusion in the absence of any additional potential $U(x)$, has been shown to be 
{\it universal} within the FFPE description 
also for arbitrary tilted washboard
potentials and temperature \cite{GH06,H06}. Recently, this astounding fact has been
related to the universal fluctuations
of anomalous mobility and weak ergodicity breaking \cite{Sokolov}. 
Ultimately, this is just 
the property of the stochastic clock and it reflects the scaling between the variance
and the mean number of stochastic periods passed within the external observed time $t$.
Surprisingly, the viscoelastic GLE subdiffusion also exhibits a {\it universal}
asymptotical scaling in tilted washboard potentials. In the $t\to\infty $ limit
it is the {\it same} as in Eq. (\ref{normal}).
Astonishingly, it works both for a vanishingly small $F$, and for an arbitrary
strong bias. Moreover, both the diffusion and drift in the
tilted washboard potentials
do not depend {\it asymptotically} on the amplitude and the form of the periodic potential 
in the case of GLE subdiffusion and are given by 
Eqs. (\ref{variance}) and (\ref{mean}), correspondingly. 
This again is very much different from the FFPE case, where Eq. (\ref{normal})
can be used only to calculate the anomalous flux response at a 
vanishingly small $F$ from the equilibrium $\langle \delta x^2(t) \rangle_{F=0}$
at $F=0$.  Also, given $\langle \delta x(t)\rangle$ at $F\neq 0$
one can calculate $\langle \delta x^2(t) \rangle_{F=0}$ using Eq. (\ref{normal}) 
and the corresponding 
subdiffusion coefficient in periodic potentials in the limit $F\to 0$,
for details see in the work\cite{H07} and below.

\section{Other similarities}
    
One more similarity emerges for the relaxation of mean fluctuation from
equilibrium in harmonic potentials, $U(x)=k x^2/2$.  Then, both the FFPE
approach and the FLE approach (in the limit $m\to 0$) yield the same
relaxation law \cite{Kou,Goychuk07a,Metzler99}, 
$\langle \delta x(t)\rangle = 
\langle \delta x(0)\rangle E_{\alpha}[-(t/\tau_r)^{\alpha}]$ with the
ultraslow position relaxation time constant $\tau_r=(\eta_{\alpha}/k)^{1/\alpha}$. 
Asymptotically, this relaxation follows a power-law, 
$\langle \delta x(t)\rangle \propto t^{-\alpha}$. 

The asymptotic 
distributions of the residence times within a half-infinite spatial 
domain (or the first 
return times to the origin in the infinite domain)
in the case of free subdiffusion
are also similar, following the same scaling law \cite{Ding,Taloni,GH04} 
$\Psi(\tau) \propto 1/\tau^{2-\alpha/2}$.
However, here the similarities end. The asymptotics for a finite-size domain
cannot be same. In particular, the mean residence time 
in any finite-size domain within the subdiffusive GLE description is finite 
\cite{Goychuk09},
whereas within the FFPE description is not, except for the case of 
injection of diffusing particles on the normal radiative boundary, where they
can be immediately absorbed\cite{GH04}. Moreover, the GLE (for arbitrary $\eta(t)$, 
including FLE) describe a Gaussian process for constant and linear forces $f(x)$
\footnote{This is just by the linearity of the transformation from the
Gaussian noise $\xi(t)$ to the stochastic process $x(t)$ as
described by Eqs. (\ref{fle}) and (\ref{gle}).}, whereas the FFPE does not correspond
to a Gaussian process in these cases, see
in Ref.\cite{Metzler}.

\section{Diffusion and transport in washboard potentials}

Let us proceed with the case of washboard potentials, where the differences
between the two discussed approaches to subdiffusion become particularly transparent.
We consider the tilted potential $U(x)=V(x)-xF$, where $V(x+L)=V(x)$ is a periodic
potential with the spatial period $L$.

\subsection{FFPE dynamics} 

In this case, one can find exact analytical results for the ensemble-averaged
nonlinear mobility $\mu_{\alpha}(F)$ using Eq. (\ref{mean}) asymptotically
also in washboard potentials. First, one finds the exact analytical expression
for the ensemble-averaged subvelocity $v_{\alpha}(F)=\mu_{\alpha}(F)F$.
The FFPE in the form (\ref{FFPE2}) is more convenient for this purpose. 
Indeed, it has the form of a fractional-time continuity 
equation with the flux $J(x)$. For the sake of
generality we consider its further generalization with a spatially-dependent
subdiffusion coefficient $\kappa_{\alpha}(x)$, 
\begin{eqnarray}
J(x,t)=-\kappa_{\alpha}(x)
e^{-\beta U(x)} \frac{\partial}{\partial x} \, e^{\beta
U(x)} P(x,t)
\end{eqnarray}
which is assumed to be periodic with the same period $\kappa_{\alpha}(x+L)
=\kappa_{\alpha}(x)$, and the generalized Einstein relation
is fullfield locally at any $x$, $\kappa_{\alpha}(x)=k_BT/\eta_{\alpha}(x)$. 
We proceed similarly to the case of normal 
diffusion \cite{Stratonovich,ReimannRev,HM09},
$\alpha=1$. A spatial period averaged density $\hat P(x,t)=\sum_{k=-M}^{M} P(x+kL,t)/(2M+1)$
should attain a steady-state regime (corresponding to a non-equilibrium
steady state for $F\neq 0$) in the limit $M\to\infty$, $t\to\infty$ 
and that becomes periodic with the period $L$, $\hat P_{\rm st}(x+L)=\hat P_{\rm st}(x)$.
The corresponding subdiffusive flux $\hat J(x)$, defined with $\hat P_{\rm st}(x)$, 
becomes a constant $J_\alpha$ in the steady state: 
\begin{eqnarray}\label{1}
J_\alpha=-\kappa_{\alpha}(x)
e^{-\beta U(x)} \frac{d}{d x} \, e^{\beta
U(x)} \hat P_{\rm st}(x)\;.
\end{eqnarray}
Then, the dynamics of the averaged mean displacement follows as 
\begin{eqnarray}\label{aux}
\sideset{_{0}}{_{*}}{\mathop{D}^{\alpha}}\langle x(t)\rangle =LJ_\alpha\;,
\end{eqnarray}
which can be shown akin to the normal diffusion case\cite{ReimannRev}. The appearance of the fractional
Caputo time derivative in the lhs of Eq. (\ref{aux}) is the only mathematical
difference as compared with the normal diffusion case.
The solution of (\ref{aux}) yields for the mean excursion
\begin{eqnarray}
\langle x(t)\rangle = v_{\alpha}^{({\rm wb})}(F) t^{\alpha}/\Gamma(1+\alpha)\;,
\end{eqnarray}
with  $v_{\alpha}^{({\rm wb})}(F)=LJ_\alpha $ 
being the subvelocity in the washboard potential.

One finds $J_{\alpha}$ and $v_{\alpha}^{({\rm wb})}(F)$ by multiplying 
Eq. (\ref{1}) with
$e^{\beta U(x)}/\kappa_{\alpha}(x)$ and integrating the result within one
spatial period. Taking into account the spatial periodicity of 
$V(x)$ and $\kappa_{\alpha}(x)$ this yields:
\begin{eqnarray}\label{2}
J_\alpha\int_y^{y+L}\frac{e^{\beta U(x)}}{\kappa_{\alpha}(x)}dx&& = -e^{\beta
U(y+L)} \hat P_{\rm st}(y+L)+e^{\beta
U(y)} \hat P_{\rm st}(y)\nonumber \\
&& =(1-e^{-\beta FL})e^{\beta U(y)}\hat P_{\rm st}(y) \;.
\end{eqnarray}
Next, multiplying (\ref{2}) with $e^{-\beta U(y)}$, integrating over $y$ within
$[0,L]$, and using the normalization $\int_{0}^L\hat P_{\rm st}(y)dy=1$ one finds 
the main result
\begin{eqnarray}\label{result}
v_\alpha^{({\rm wb})}(F)=\frac{(1-e^{-\beta FL})L}{
\int_0^L e^{-\beta U(y)}dy\int_y^{y+L}
\frac{e^{\beta U(x)}}{\kappa_{\alpha}(x)}dx}\;.
\end{eqnarray}
Accordingly, the nonlinear anomalous mobility is $\mu^{(\rm wb)}_{\alpha}(F)=
v_\alpha^{({\rm wb})}(F)/F$. This presents a further generalization of
the result for subvelocity in Refs. \cite{GH06,H06} 
to a spatially-dependent subdiffusion coefficient $\kappa_{\alpha}(x)$. 
The subdiffusion coefficient in the 
unbiased washboard potential for $F=0$ can also be found using
the generalized Einstein relation $\kappa^{(\rm wb)}_{\alpha}(F=0)=k_BT 
\mu^{(\rm wb)}_{\alpha}(F=0)$. It reads,
\begin{eqnarray}\label{result2}
\kappa_\alpha^{({\rm wb})}(F=0)=\frac{L^2}{
\int_0^L e^{-\beta V(y)}dy\int_0^{L}
\frac{e^{\beta V(x)}}{\kappa_{\alpha}(x)}dx}\;,
\end{eqnarray}
and for $\kappa_{\alpha}=const$ this is the result  of the work\cite{H07}.
For constant $\kappa_{\alpha}$ and a number of different potentials $V(x)$, 
temperatures $T$
and biasing forces $F$, 
these two general results were beautifully confirmed by numerical simulations
of the underlying CTRW \cite{GH06,H06,H07} on a lattice from which the FFPE in the form 
(\ref{FFPE2}) was derived in the work\cite{GH06}. These simulations also confirmed 
the universality of the scaling relation (\ref{anomal}) within the FFPE approach. 
Surprisingly, it remains invariant
also in the presence of a driving which is periodic in time, 
in the biased case $F\neq 0$ \cite{H09}, featuring thus the universality class of
subdiffusion governed by a stochastic clock with divergent mean period
and characterized by the only parameter $\alpha$. The above $v_{\alpha}^{(\rm wb)}$
is the ensemble-averaged result. The subvelocities of individual particles
remain randomly distributed in the limit $t\to\infty$ and 
they follow a universal subvelocity distribution which reflects the distribution of
random individual time of travelling particles, as it has been clarified in
Ref. \cite{Sokolov}.  Both the weak ergodicity breaking and the universal fluctuations of
anomalous mobility within the FFPE approach are 
ultimately related to this remarkable
property of the stochastic time.

\subsection{GLE dynamics in periodic potentials}

The GLE subdiffusion distinctly differs in the physical mechanism and
this leads to quite different results for washboard potentials\cite{Goychuk09,G10}. 
First of
all, it is  {\it asymptotically} ergodic and self-averaging over a single
trajectory yields a quite  definite non-random result \cite{Goychuk09}. 
%\footnote{For the transport over several neighboring potential
%wells no self-averaging occurs. Such a short-range transport
%remains nonergodic.}. 
No additional
ensemble averaging is required. Moreover, it turns out that both the
particle anomalous mobility $\mu^{(\rm wb)}_{\alpha}$  and the
subdiffusion coefficient  $\kappa^{(\rm wb)}_{\alpha}$ do not
depend {\it asymptotically} neither on the potential $V(x)$, nor on the
bias $F$ being universal and the same as for biased GLE subdiffusion
in the absence of periodic potential, obeying the generalized Einstein
relation. The transition to this asymptotic regime is, however, very
slow and it strongly depends on the amplitude of the periodic
potential $V_0$ and  the temperature $T$. Because of this slowness of the 
transient aging,  this asymptotic regime
will not necessarily be relevant on a finite time scale 
for anomalous transport in finite-size systems. This is  especially so if the periodic
potential amplitude exceeds the thermal energy by many times. However, this
remarkable property features the very mechanism of the GLE subdiffusion, which
is based on the long-range velocity and displacement correlations and not on
diverging mean residence time within a potential well, in clear contrast to
the CTRW subdiffusion with independent increments. 
It outlines a quite different universality class
of subdiffusion. This is the long-range anti-persistence
which limits asymptotically the GLE subdiffusion and transport processes in the 
washboard potentials. Since the mean residence time in a potential well 
is finite \cite{Goychuk09}, a coarse graining over the potential period, 
which makes the sojourns in the
trapping potential wells irrelevant, becomes {\it asymptotically} possible.
In fact, upon increasing the potential height the escape kinetics out 
of a potential well (being asymptotically stretched-exponential) becomes ever closer
to the normal exponential kinetics \cite{Goychuk09}, where it becomes  described 
by the non-Markovian rate theory\cite{HTB90,HM82}. 
This does not mean, however, that the diffusion spreading over many spatial periods 
becomes normal. As a matter of fact, in the unbiased periodic potentials the
diffusion cannot become faster than the free subdiffusion and this is a reason why 
the asymptotic limit of free subdiffusion is attained.  A signature of this 
universality has been revealed theoretically for quantum transport in sinusoidal
potentials for the case of sub-Ohmic thermal bath which classically corresponds
to the considered case of fractional sub-diffusive friction. Technically this
was done by
using two different approaches, one perturbative\cite{Chen} and one non-perturbative
based on a quantum duality transformation between the quantum dissipative washboard
dynamics coupled to a sub-Ohmic bath and a quantum dissipative tight-binding dynamics
coupled to a super-Ohmic bath\cite{WeissBook}. In the quantum case, there are 
also tunneling
processes which are accounted for. Our numerical results for the classical
Brownian dynamics indicate, however, that this feature is purely 
classical and, moreover, it is universal, i.e. is beyond the particular case of
sinusoidal potentials\cite{G10}.  
It is {\it not} caused by the quantum-mechanical effects.

Our numerical simulation approach is also insightful and it can be considered
as an independent theoretical route to model anomalous diffusion and transport processes.
The idea is to approximate the non-Markovian GLE dynamics with a power-law kernel
by a {\it finite-dimensional} Markovian dynamics of a sufficiently high 
dimensionality $D$ \cite{Marchesoni,Kupferman,Goychuk09,Siegle}. Here,
"sufficient" means the following: having subdiffusion 
extending over $r$ time-decades one finds a $D$-dimensional Markovian dynamics
whose projection on the $(x,v)$ plane approximates the GLE dynamics over the required
time range within the
accuracy of stochastic simulations, as it can be checked for the cases where
an exact solution of the GLE dynamics is available 
(free or  biased subdiffusion, subdiffusion
in harmonic potentials). Increasing $D$ one can cover larger $r$ of experimental
interest and the embedding
dimension $D$ turns out to be finite 
to arrive at the asymptotic results
valid for the strict power law kernel. Needless to say that the practically
observed cases of anomalous diffusion hardly extend over more than 6 time decades
(typically several only) which underpins the practical value of our approach.
 
We expand the power law kernel into a sum of exponentials
\begin{eqnarray}\label{expansion}
\eta(t)=\frac{\eta_{\alpha}}{\Gamma(1-\alpha)}C_{\alpha}(b)
\sum_{i=1}^{N}\nu_i^\alpha \exp(-\nu_i t)
\end{eqnarray}
obeying a fractal scaling with $\nu_i=\nu_0/b^{i-1}$, where $b>1$ is a scaling 
parameter, $\nu_0>0$ is high-frequency (short-time) cutoff corresponding to
the fastest time scale $\tau_0=1/\nu_0$ in the hierarchy of the 
relaxation time constants,
$\tau_i=\tau_0b^{i-1}$, of viscoelastic memory kernel. 
$C_{\alpha}(b)$ is  a numerical constant to provide
a best fit to $\eta(t)=\eta_{\alpha}t^{-\alpha}/\Gamma(1-\alpha)$ in the
interval $[\tau_0,\tau_0 b^{N-1}]$. In the theory of anomalous relaxation
similar expansions are well-known \cite{Hughes,Palmer}. In the present context, 
the approach corresponds to an approximation of the
fractional Gaussian noise by a sum of uncorrelated 
Ornstein-Uhlenbeck (OU) noises, 
$\xi(t)=\sum_{i=1}^{N}\zeta_i(t)$, with 
autocorrelation functions, $\langle\zeta_i(t)\zeta_j(t')\rangle=k_BT \kappa_i
\delta_{ij}\exp(-\nu_i|t-t'|)$. This idea is also known in the theory of
$1/f$ noise \cite{Weissman}. 
For $t>\tau_0 b^{N-1}$ the tail of (\ref{expansion}) is 
exponential and the diffusion becomes normal for $t\gg \tau_0 b^{N-1}$.
However, by increasing $N$ one can enlarge the corresponding time scale
and even make it practically irrelevant.
The subdiffusion can be modelled in this way over $r=N\log_{10} b-2$ time decades
and the corresponding embedding dimension, $D=N+2$, can be rather small. Such fits
are known to exhibit logarithmic oscillations superimposed on the power law \cite{Hughes}.
However, their amplitude can be made negligibly small if to choose $b$
sufficiently small, e.g. for $b=2$ they become already barely detectable. 
Nevertheless, even the decade scaling with $b=10$
suffices to arrive at excellent (within the statistical errors of Monte Carlo
simulations) approximation of the FLE dynamics by a {\it finite-dimensional} Markovian
dynamics over a huge range of time scales.
Weak logarithmic sensitivity of $r$ to $b$  and linear dependence  on $N$ allows one 
to improve the quality of
Markovian embedding at a moderate computational price. The choice of 
Markovian embedding which corresponds to (\ref{expansion}) is not 
unique\cite{Kupferman,Siegle}. 
A particular one is the following \cite{Goychuk09}: 
\begin{eqnarray}\label{embedding1}
\dot x&=& v\;,\nonumber \\
m\dot v & =& f(x,t)+
\sum_{i=1}^{N}u_i(t) \;,\nonumber \\
\dot u_i& = &-k_i v-\nu_iu_i+\sqrt{2\nu_i k_i k_BT}\xi_i(t) \;,
 \end{eqnarray}
where $k_i=C_\alpha(b)\eta_\alpha\nu_i^\alpha/\Gamma(1-\alpha)>0$ and 
$\xi_i(t)$ are independent unbiased white Gaussian noise sources, 
$\langle \xi_i(t)\xi_j(t')\rangle=\delta_{ij}\delta(t-t')$. 
Indeed, integrating out the auxiliary force variables $u_i$ in Eq. (\ref{embedding1})
it follows that the resulting dynamics is equivalent to
the GLE (\ref{gle}), (\ref{fdt}) with the kernel (\ref{expansion}), provided that $u_i(0)$ are unbiased
random Gaussian variables with variances $\langle u_i^2(0)\rangle=k_ik_BT$.
The latter condition ensures the stationarity of $\xi(t)$ in the GLE (\ref{gle}),
as well as validity of the FDR (\ref{fdt}) {\it for all times}.
Using different non-thermal preparations of $u_i(0)$ one can study the influence
of {\it initial} non-stationarity of the noise $\xi(t)$ in the GLE 
on the Brownian dynamics\cite{Siegle}. In this aspect, our approach is even more flexible and 
more general than the standard GLE approach.

The auxiliary variables $u_i$
can be interpreted as elastic forces, $u_i=-k_i(x-x_i)$, exerted 
by some overdamped particles 
with positions $x_i$, which are coupled to the central Brownian particle with elastic
spring constants $k_i$ and are subjected to viscous friction with frictional
constants $\eta_i=k_i/\nu_i=C_{\alpha}(b)
\eta_{\alpha}\tau_i^{1-\alpha}/\Gamma(1-\alpha)$ 
and the thermal random forces of environment. This corresponds to
motion of $N+1$ particles in a potential $U(x,\{ x_i\})=U(x,t)+(1/2)
\sum_{i=1}^N k_i(x-x_i)^2$. The Brownian particle is massive (inertial effects
are generally included), all other ``particles'' are overdamped 
(massless, $m_i\to 0$). For example, one can imagine 
that some coordination spheres
of the viscoelastic environment stick to the Brownian particle and are co-moving.
Their influence can be effectively represented by $N$ ``quasi-particles''.  
In this insightful physical interpretation, 
our embedding scheme is equivalent to: 
\begin{eqnarray}\label{embedding2}
m\ddot x & =& f(x,t)-
\sum_{i=1}^{N}k_i(x-x_i) \;,\nonumber \\
\eta_i\dot x_i& = &k_i (x-x_i)+\sqrt{2\eta_ik_BT}\xi_i(t) \;.
 \end{eqnarray}
It worth to notice that in this approach the mass of the Brownian particle
and therefore the inertial effects are important. In order to perform an overdamped
limit $m\to 0$, one has to include the viscous frictional force $-\eta_0 \dot x$ 
acting directly on 
the particle and the corresponding random force. Then, in the limit $m\to 0$,
one obtains
\begin{eqnarray}\label{embedding3}
\eta_0 \dot x & =& f(x,t)-
\sum_{i=1}^{N}k_i(x-x_i)+ \sqrt{2\eta_0k_BT}\xi_0(t) \;,\nonumber \\
\eta_i\dot x_i& = &k_i (x-x_i)+\sqrt{2\eta_ik_BT}\xi_i(t) \;,
 \end{eqnarray} 
where $\xi_0(t)$ is a zero-mean Gaussian random force of unit intensity which is not
correlated with the set $\{ \xi_i(t) \}$. 
However, it was noticed \cite{Burov} that the inertial effects are
generally important for the subdiffusive GLE dynamics and therefore we take them
into account. Of  course, here emerges one more difference with the alternative 
description of subdiffusion within the FFPE (\ref{FFPE1}), (\ref{FFPE2}). 
 
\begin{figure}
\includegraphics [width=0.5\linewidth]{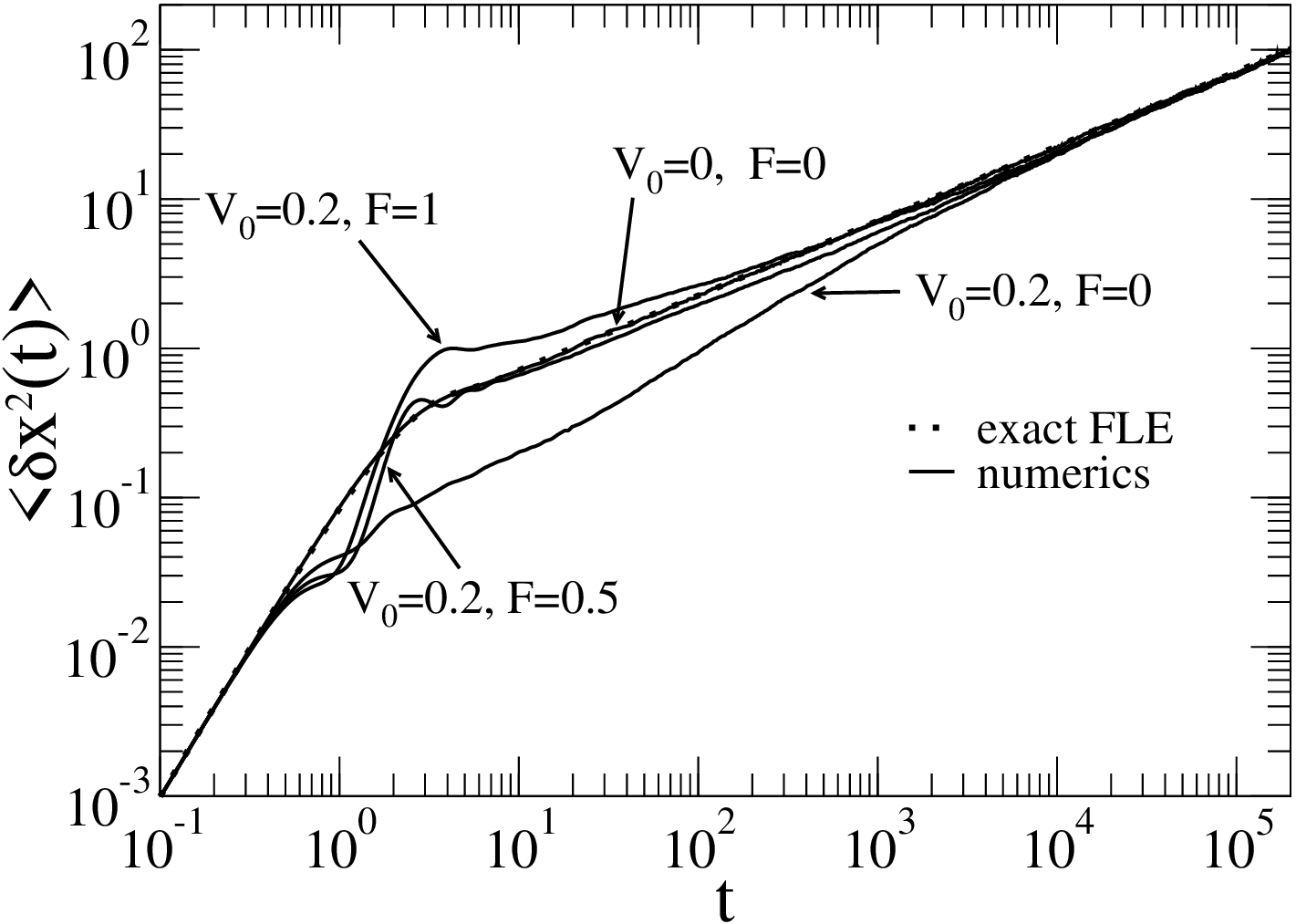}
\caption{Anomalous diffusion in the potential $U(x)=-V_0\sin(2\pi x/L)-Fx$ for various $V_0$
and $F$ at $T=0.1$ for $\alpha=0.5$. Notice an excellent agreement (differences
practically cannot be detected in this plot) 
of simulations
with the exact FLE result, $\langle \delta x^2(t)\rangle
=2v_T^2t^2E_{2-\alpha,3}[-(t/\tau_v)^{2-\alpha}] $, 
in the absence of periodic potential $V_0=0$. Scaling: time in 
$\tau_v=(m/\eta_{\alpha})^{1/(2-\alpha)}$, distance in $L$, energy in 
$m(L/\tau_v)^2$, force in $mL/\tau_v^2$ and temperature in 
$mL^2/(\tau_v^2k_B)$.}
\label{Fig2}
\end{figure} 
 
\begin{figure}
\vspace{0.5cm}
\includegraphics [width=0.5\linewidth]{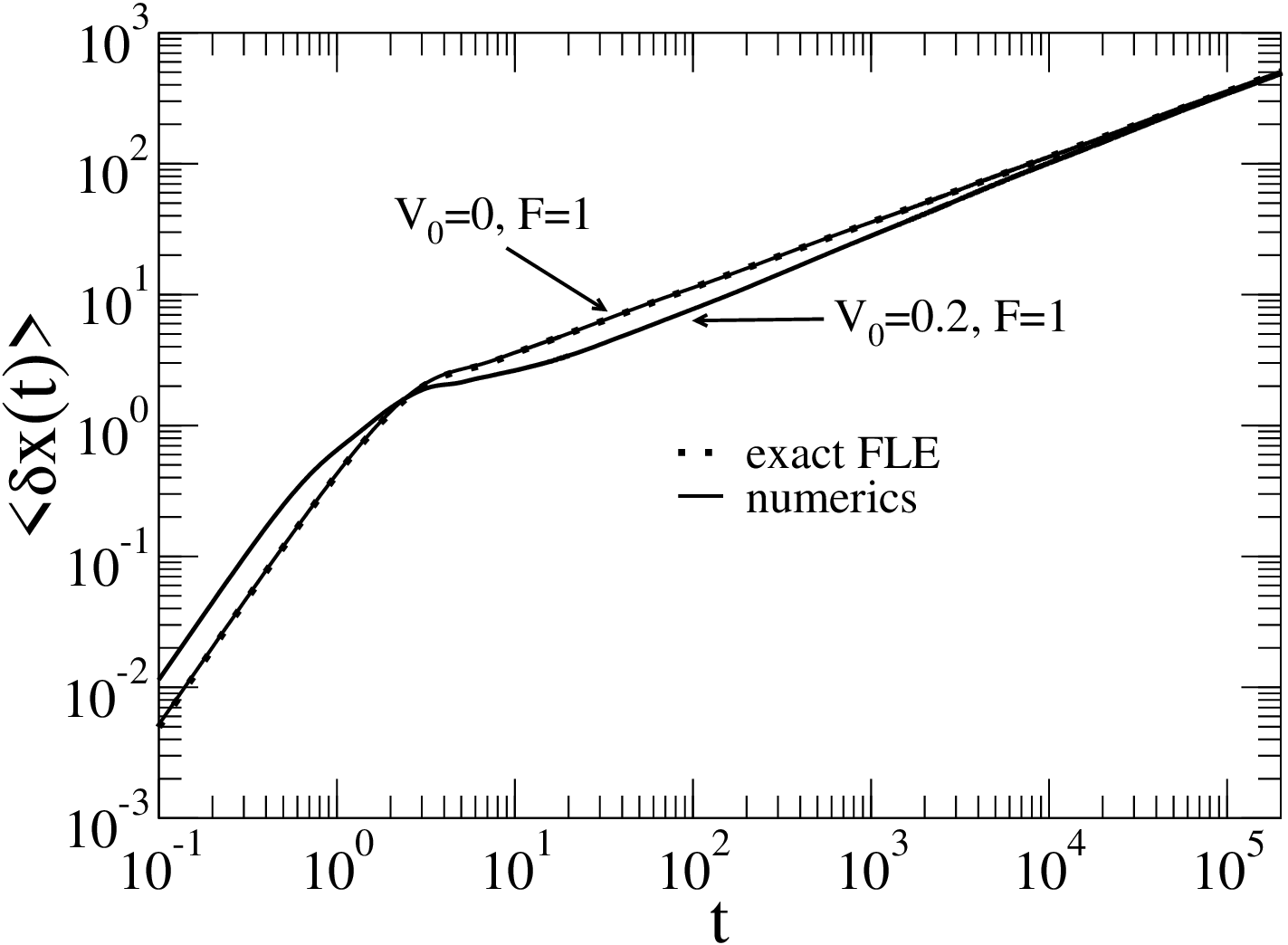}
\caption{Anomalous transport for $F>0$, see Fig.\ref{Fig2} for details.}
\label{Fig3}
\end{figure}

A proper fractal scaling of coefficients $k_i$ and $\eta_i$ with $i$ (see above) 
allows one to 
model viscoelastic subdiffusion
over arbitrary time scales of the experimental interest. 
One can numerically solve these stochastic differential equations (\ref{embedding1})
e.g. with a standard stochastic Heun method\cite{Gard} (second order Runge-Kutta method)
as done in Refs. \cite{Goychuk09,G10}. 
An example of such simulations is given in Figs. \ref{Fig2}, 
\ref{Fig3} for $\alpha=0.5$, $\nu_0=100$, $b=10$, $C_{\alpha}(b)=1.3$, 
$N=12$ and $k_B T=0.1$. The following scaling is used: 
time in the units of $\tau_v$ \footnote{This is a natural 
scaling of the velocity autocorrelation function in time. 
Other scalings are but also possible\cite{Goychuk09,G10}. 
They are more suitable to consider dynamical regimes close to overdamped.},
distance in the units of $L$. All the energy units are then scaled in 
$\Delta E=m(L/\tau_v)^2$ and the force units in $mL/\tau_v^2$.
Stochastic Heun method is used to integrate Eq. (\ref{embedding1}) with
a time step $\Delta t=(1-5)\cdot 10^{-3}$ until $t_{\rm max}=2\cdot 10^5$ 
and $n=10^4$ trajectories are used for
the ensemble averaging. Stochastic numerics are compared against the exact
results for the free subdiffusion and  for the mean displacement 
under a constant biasing force. The agreement is excellent. The considered
particular  embedding
still works as an approximation to the FLE dynamics until $t=10^8$. If one needs to
describe subdiffusion on an even longer time scale one can increase $N$. If one needs
a better precision of approximation one can make $b$ smaller. 
Initially all the particles are localized at the origin, $x=0$, with the velocities
thermally distributed at the temperature $T$. 
For the time span 
$t\lesssim\tau_v$ the motion
is always ballistically persistent (superdiffusion). This reflects the
inertia of the Brownian particle. It assumes the subdiffusive 
character for $t\gg \tau_v$, when the VACF
is negative. The presence of a periodic potential
$V(x)=-V_0\sin(2\pi x/L)$ dramatically changes
both subdiffusion, $\langle x^2(t)\rangle-\langle x(t)\rangle^2$,
as well as subdiffusive transport, $\langle x(t)\rangle$, on  intermediate time
scale. However, the long time asymptotics
of free or biased subdiffusion are  gradually attained. 
The initial behavior still within one
potential well remains ballistic.  One can 
conclude that both subdiffusion and subdiffusive transport are indeed asymptotically
insensitive to the presence of periodic potential within the GLE approach. 
This finding is in
a striking contrast with the FFPE approach. 
However, the transient to
this asymptotic regime can be very slow, depending on the amplitude of the periodic
potential and temperature. 

An interesting phenomenon is also accelerated subdiffusion occurring on 
an intermediate time scale in tilted washboard potentials, as compared with the 
free subdiffusion. It can be detected in Fig. \ref{Fig2} for a strong yet subcritical
bias $F=1<F_{\rm cr}=2\pi V_0/L\approx 1.2566...$. This calls to 
mind  the acceleration of normal diffusion in tilted 
washboard potentials\cite{Reimann3}. However, this accelerated
subdiffusive phenomenon
occurs only on an intermediate time scale because  asymptotically the GLE 
subdiffusion is not sensitive to the presence of the potential.
One more interesting effect occurs for the initially ballistic
transport. It first seems paradoxical that in the trapping potential 
the initial transport becomes  
faster than in the absence of potential 
and not vice versa, see in Fig. \ref{Fig3}. The result 
can be understood in view of the fact that
the minimum of the potential under the strong bias $F$ is essentially
displaced in the direction of biasing force and the particles are {\it initially} 
accelerated 
by the additional to $F$ force  stemming from the periodic potential. \\ \vspace{0.5cm}

%\vspace{3cm}

\section{Summary and conclusions}

With this Chapter, we reviewed and scrutinized 
two different approaches, the FFPE approach and the FLE approach, to anomalously 
slow diffusion and transport in nonlinear force fields with a focus
on applications  in tilted periodic 
potentials. In spite of some similarities in the case of constant or linear forcings
it was shown that the nonlinear dynamics radically differ, obeying
asymptotically two different universality classes. A first one reflects the
universal fluctuations of intrinsic time clock and is closely tight
to a weak ergodicity breaking. 
In contrast, within the GLE and FLE approach the long-range 
antipersistence of the velocity and position fluctuations renders the asymptotic dynamics
ergodic. One approach seems more appropriate for the disordered solids,
or glass-forming liquids below the glass-forming 
transition, as characterized by the nonergodic 
glass phase. Another one seems more appropriate for the regime 
above but close to the glass
transition, or for crowded viscoelastic environments like cytosols in  biological
cells. We have left out further pronounced differences between the 
FFPE and FLE approaches in the
case of time-dependent fields\cite{SK06,HPRL07,G07,H09,West2}. 
We are confident that  
our results not only shed light on the origin of profound differences, but also
will stimulate a further development of both approaches to subdiffusion, and
possibly other interrelationships emerging in random potentials.

\acknowledgements

We would like to thank  E. Heinsalu, M. Patriarca, G. Schmid, and P. Siegle for
a very fruitful collaboration on anomalous transport in washboard potentials. 
This work was supported by the Deutsche Forschungsmeinschaft, grant 
No. GO 2052/1-1 (I.G.) and through Nanosystems Initiative Munich (P.H.).

\appendix

\section{Continuous time random walk and random clock}

Consider a lattice with period $a$ and a particle jumping with probabilities 
$q_+$ and $q_-$, $q_++q_-=1$, to the neighboring sites after a random clock
characterized by the residence time distribution (RTD) 
$\psi(\tau)$ ``ticked'' on the next jump.
Within the physical time interval $t$ there will be a variable random number
of intrinsic time periods $n(t)$. The probability to make $m$ steps forward 
and $n-m$ steps backward  after $n$ periods is given by the binomial 
distribution, $P(m,n)=n!/[m!(n-m)!]q_+^mq_-^{n-m}$. Using it one can 
calculate
the first two moments, $\langle x^k\rangle=a^k\sum_{m=0}^n (2m-n)^k P(m,n), k=1,2$, 
of the particle displacement after $n$ periods:
\begin{eqnarray}  
\langle x(t)\rangle & = & a(q_+-q_-)n(t) \nonumber \\
\langle x^2(t)\rangle & = & a^2\left [n^2(t)(q_+-q_-)^2+ 4n(t)q_+q_- \right]\;.
\end{eqnarray}
They are still random quantities because of the randomness of $n(t)$. Each particle
has an individual number of periods completed until $t$. For the additional 
ensemble average one obtains 
\begin{eqnarray}  
\langle \langle x(t)\rangle \rangle_{\rm ens} & = & a(q_+-q_-)
\langle n(t)\rangle_{\rm ens} \nonumber \\
\langle \langle x^2(t)\rangle\rangle_{\rm ens} & = & a^2\left 
[\langle n^2(t)\rangle_{\rm ens} (q_+-q_-)^2+ 
4\langle n(t)\rangle_{\rm ens} q_+q_- \right]\;
\end{eqnarray}
and for the ensemble-averaged variance 
$\langle \langle [x-\langle \langle x\rangle\rangle_{\rm ens}]^2
\rangle\rangle_{\rm ens}=\langle \langle x^2\rangle\rangle_{\rm ens} -
\langle \langle x\rangle\rangle^2_{\rm ens}$
\begin{eqnarray}  
\langle \langle \delta x^2(t)\rangle\rangle_{\rm ens} & = & a^2
\left [\langle \delta n^2(t)\rangle_{\rm ens} (q_+-q_-)^2+ 
4\langle n(t)\rangle_{\rm ens} q_+q_- \right]\; 
\end{eqnarray}
where 
$\langle \delta n^2(t)\rangle_{\rm ens}=\langle n^2(t)\rangle_{\rm ens}-
\langle n(t)\rangle_{\rm ens}^2$ is the variance of random periods passed.
Notice
that $\langle \langle x^2\rangle\rangle_{\rm ens} -
\langle \langle x\rangle\rangle^2_{\rm ens}\neq 
\langle \langle [x-\langle x\rangle]^2
\rangle\rangle_{\rm ens}$.
Clearly, for a regular clock, $\langle \delta n^2(t)\rangle_{\rm ens}=0$ and 
the corresponding contribution to the ensemble-averaged position variance
is absent. To simplify the notations, we further denote the ensemble averages as
$\langle ...\rangle$ rather than $\langle\langle ...\rangle\rangle_{\rm ens}$.

The physical time $t$ can be measured by the sum of {\it independent} 
stochastic periods $\tau_k$ already completed and one not yet completed
period $\tau_{n+1}^*$, $t=\sum_{k=1}^{n(t)}\tau_k+\tau_{n+1}^*$, with
$n=0,1,2,...,\infty$. Therefore,
the probability distribution $p(n,t)$ to have $n$ time periods within $t$,
$\sum_{n=0}^{\infty}p(n,t)=1$,
is the $n+1$-time convolution of the RTDs $\psi(\tau)$ ($n$ times) and of the survival 
probability $\Phi(\tau)=\int_{\tau}^{\infty}\psi(\tau)d\tau$. 
Its Laplace-transform reads
\begin{eqnarray}
\tilde p(n,s)=\frac{1-\tilde \psi(s)}{s}[\tilde \psi(s)]^n
\end{eqnarray}
in terms of the Laplace-transformed $\psi(\tau)$. Let's consider 
$\tilde \psi(s)\approx 1-(s\tau_{\rm sc})^{\alpha}$ for $s\tau_{\rm sc}\to 0$, 
where
$\tau_{\rm sc}$ is a time unit of measurements. The continuous spatial limit is achieved
when $a\to 0$, $\tau_{\rm sc}\to 0$ with $\kappa_{\alpha}=
a^2/\tau_{\rm sc}^{\alpha}$ being
a constant. For a finite $\tau_{\rm sc}$, considering the scaling limit $n\to\infty$, 
$s\tau_{\rm sc}\to 0$
with $n(s\tau_{\rm sc})^\alpha$ being finite, one obtains
\begin{eqnarray}\label{rtime}
\tilde p(n,s)=\tau_{\rm sc}(s\tau_{\rm sc})^{\alpha-1}\exp[-n(s\tau_{\rm sc})^\alpha],
\end{eqnarray} 
where $n$ is considered as a continuous variable and 
\begin{eqnarray}
\tau(t)=n(t)\tau_{\rm sc}
\end{eqnarray}
is the intrinsic random time. Notice that for $\alpha=1$ one
finds, $p(n,t)=\delta(n-t/\tau_{\rm sc})$ by inversion 
to the time domain. That means to say that $\tau(t)=t$ is not random. 
For $0<\alpha<1$,
$p(n,t)$ can be expressed  via the one-sided Levy distribution
density ${\cal L}_{\alpha}(t)$ whose Laplace transform reads 
$\tilde {\cal L}_{\alpha}(s)=\exp(-s^\alpha)$. Then,
all the moments $\langle n^k(t)\rangle$ can be easily found from
(\ref{rtime}) to read 
\begin{eqnarray}
\langle n^k(t)\rangle=\frac{\Gamma(1+k)}{\Gamma(1+k\alpha)}(t/\tau_{\rm sc})^{k\alpha}.
\end{eqnarray} 
In spite of the fact that the mean time interval $\langle \tau\rangle$ 
does not exist all the moments of the 
intrinsic time $\tau(t)$ are finite. 
This might seem paradoxical. However, the intrinsic
time scales with the number of stochastic periods passed and if the mean period
does not exist the moments of $n(t)$ are nevertheless finite for any finite $t$.
This is because  a frequent occurrence of very long stochastic time periods 
within some fixed $t$ 
implies a {\it smaller} value of $n(t)$. 
In particular,
\begin{eqnarray}
\langle n(t)\rangle & = &(t/\tau_{\rm sc})^{\alpha}/\Gamma(1+\alpha),\nonumber \\
\langle n^2(t)\rangle &=&2(t/\tau_{\rm sc})^{2\alpha}/\Gamma(1+2\alpha) ,\;
\end{eqnarray} 
and 
\begin{eqnarray}
\frac{\langle \delta n^2(t)\rangle}{\langle n(t)\rangle^2}=
\frac{\langle \delta \tau^2(t)\rangle}{\langle \tau(t)\rangle^2}=
\frac{2\Gamma^2(1+\alpha)}{\Gamma(1+2\alpha)}-1
\end{eqnarray}  
is the most important property of the stochastic clock. 
It is primarily responsible for the
discussed universality class of the CTRW-based subdiffusion associated with 
the universal fluctuations, 
and the  weak ergodicity breaking.

%\bibliography{ws-rv-sample}

%\printindex[aindx]                 % to print author index
                      % to print subject index
\end{document}